\documentclass[aps,pra,superscriptaddress,twocolumn,showpacs]{revtex4}
\usepackage[colorlinks, linkcolor=blue, citecolor=blue, urlcolor=blue, breaklinks=true]{hyperref}
\usepackage{amsmath}
\usepackage{graphicx}
\usepackage{amssymb}
\usepackage{bbm, color}

\newcommand*{\cl}[1]{{\mathcal{#1}}}
\newcommand*{\bb}[1]{{\mathbb{#1}}}
\newcommand{\ket}[1]{\left|#1\right>}

\newcommand{\proj}[2]{| #1 \rangle\!\langle #2 |}
\newcommand*{\tn}[1]{{\textnormal{#1}}}
\newcommand*{\1}{{\mathbbm{1}}}

\begin{document}

\title{Correspondence between maximally entangled states in discrete and Gaussian regimes }
\author{Youngrong Lim}
\affiliation{Department of Mathematics and Research Institute for Basic Sciences, Kyung Hee University, Seoul 02447, Korea}
\author{Jaewan Kim}
\affiliation{School of Computational Sciences, Korea Institute for Advanced Study, Seoul 02455, Korea}
\author{Soojoon Lee}
\affiliation{Department of Mathematics and Research Institute for Basic Sciences, Kyung Hee University, Seoul 02447, Korea}
\affiliation{School of Computational Sciences, Korea Institute for Advanced Study, Seoul 02455, Korea}
\author{Kabgyun Jeong}
\email{kgjeong6@snu.ac.kr}
\affiliation{IMDARC, Department of Mathematical Sciences, Seoul National University, Seoul 08826, Korea}
\affiliation{School of Computational Sciences, Korea Institute for Advanced Study, Seoul 02455, Korea}

\date{\today}

\begin{abstract}
We study a general corresponding principle between discrete-variable quantum states and continuous-variable (especially, restricted on Gaussian) states via quantum purification method. In the previous work, we have already investigated an information-theoretic correspondence between the Gaussian maximally mixed states (GMMSs) and their purifications known as Gaussian maximally entangled states (GMESs) in [Phys. Lett. A {\bf 380}, 3607 (2016)].  We here compare an $N\times N$-dimensional maximally entangled state to the GMES we proposed previously, through an explicit calculation of quantum fidelity between those entangled states. By exploiting the results, we naturally conclude that our GMES is more suitable to the concept of \emph{maximally entangled} state in Gaussian quantum information, and thus it might be useful or applicable for quantum information tasks than the two-mode squeezed vacuum (TMSV) state in the Gaussian regime.
\end{abstract}

\maketitle

\section{Introduction} \label{intro}

Quantum information science in a finite dimensional Hilbert space not only in theory but also in experiment has been well established~\cite{NC00,W13} since 1980s. It enables us to break the limit which classical physics cannot overcome, for many computation and communication tasks, such as quantum speed-up algorithms~\cite{S97,G97}, quantum key distribution (QKD) protocols~\cite{BB84,E91,B92}, and non-additive channel-capacity problems~\cite{SY08,H09,LWZG09}. 

However, for the \emph{infinite} dimensional Hilbert space, there is still a subtle gap to the finite case of its concrete interpretation, because it has infinite degrees of freedom in general, and thus we must face with many unphysical situations. For example, we know that EPR state~\cite{EPR} is a kind of maximally entangled states in the continuous-variable (CV) regime, but it is actually unphysical (i.e., it is unnormalizable). Nevertheless, many investigation of CV systems are steadily performed, since all of those experiments are well-worked and quite easy in the setting of the continuous-variable regime. In particular, the usual components such as beam-splitter, squeezer, and laser in quantum optics, are well described by only Gaussian states and Gaussian unitary operations that are special kinds of continuous-variable states and operations. 

Though both discrete-variable (DV) and CV quantum information theory are important, however, they have been investigated separately without any obvious correspondence in quantum information communities. Finding a general corresponding principle between CV and DV quantum regimes was suggested as in Ref.~\cite{Caslav03}, but it is still missing a comprehensible connection. As an easier question, first we can ask a correspondence relation between DV quantum states and Gaussian states, instead of general CV states themselves. The Gaussian states are still living in the infinite Hilbert space (i.e., phase space) but all the properties we are interested in are encoded only on the covariance matrices whose dimensions are finite. Many researches have been performed on Gaussian quantum information~\cite{rmp} including maximally entangled states in Gaussian regime~\cite{GMMES}, on the contrary there is no clear connection from infinite to finite dimensional quantum states.

In the previous work~\cite{JeongLim}, we have proposed several candidates for Gaussian maximally entangled states (GMESs) given by the purification process over Gaussian maximally mixed states (GMMSs) with coherent states as well as arbitrary squeezed Gaussian states~\cite{Bradler, JKL15}. Here, we take comparisons between our specific GMES and the maximally entangled state (MES) having $N \times N$ dimension. When an appropriate parameter is chosen, the quantum fidelity between the MES ($N >200$) and our GMES is almost close to 0.99, that is much greater value than the case of two-mode squeezed vacuum (TMSV) state, which is another well-known candidate for the GMES. Also we can check Bell violation for two-qutrit cases and this result confirms the fact that our GMES shows more violation of Bell's inequality than TMSV state under the constraint of the same average-photon number. 

In Section~\ref{genco}, we introduce several ideas for the corresponding framework between DV quantum states and general quantum CV (or Gaussian) states. In Section~\ref{CVDV}, we focus on the GMES case and give a brief review of previous works including quantum purification method in the Gaussian regime. By using the analysis of Bell functions and quantum fidelity on Gaussian quantum states, main results are described in Section~\ref{main} with several remarks. Finally, we summarize our results and discuss for possible future works in Section~\ref{discussion}.

\section{Correspondence framework: quantum DV vs. CV systems} \label{genco}

Since a CV quantum state is lying on the infinite dimensional Hilbert space, it has the infinite degrees of freedom in general. Therefore, it is not an overestimation that a generic CV quantum state possesses potentially higher resources than a DV state for performing a quantum information processing. In this sense, finding a specific correspondence relation between the CV state and  the DV state with arbitrary dimension is a reasonable and important task. One of the methods is that, in advance, find a mapping between DV states having different dimensions, and take a limit of dimension going to infinity for one of those states. For an $n$-dimensional quantum state, we have generators for SU($n$) algebra from the transition operators~\cite{Hioe}. Then we consider another $N$-dimensional quantum state with $N\geq n$ and assume that $N$ can be divided by $n$ for the simplicity. By using an appropriate coarse-graining method, we can have $N/n$ of SU($n$) algebras and sum those up to finally get the generators of SU($n$) algebra fully spanned in $N$-dimension~\cite{Caslav03}. The problem is arising, however, when we take the limit $N \rightarrow \infty$ for getting CV state. In this limit, roughly speaking, a non-MES can be mapped onto the MES of finite dimension. In other words, there might be several ways of mapping MES in finite dimension to the CV state.

There is another example showing the correspondence between CV and DV states under nonlinear quantum optical settings~\cite{Van,Lee,Kim}. In fact, a coherent state is a superposition of pseudo-number states, which are $d$-modulo photon number states, and can be written as $\ket{\alpha}={1 \over \sqrt{d}} \sum_{k=0}^{d-1}\ket{k_d}$,
where $\ket{k_d}$ is a pseudo-number state with `$k \mod d$' number of photons \cite{Kim}. 
After a cross-Kerr interaction represented by $e^{{{2\pi i}\over{d}}\hat{n}_1\hat{n}_2}$ on the two-mode initial state $\ket{\alpha}_1\ket{\alpha}_2$, the MES of pseudo-number states and pseudo-phase states can be produced, if $|\alpha| > d$ holds, as follows:
\begin{equation} \label{selfkerr}
\ket{\alpha}_1\otimes\ket{\alpha}_2\overset{\underset{\tn{Cross-Kerr}}{}}{\longmapsto}{1\over\sqrt{d}}\sum_{k=0}^{d-1}\ket{k_d}_1\otimes|\tilde{k}_d\rangle_2,
\end{equation}
where $|\tilde{k}_d\rangle$ is a pseudo-phase state which is equivalent to a ${2\pi k}\over{d}$-phase-shifted coherent state $|e^{{2\pi k i}\over{d}} \alpha\rangle$.

The simplest example is the quantum information processing with even- and odd-cat states representing logical 0- and 1-qubit~\cite{Yurke,Sanders}. This is quite simple but the pseudo-number states become orthogonal only when $|\alpha|$ gets larger than $d$ and stronger cross-Kerr nonlinearity is required. Experimental generation of this type of MES from coherent states with large amplitude is quite challenging and the strength of Kerr interaction is extremely weak usually.

Although it is possible to create a maximally entangled states through the nonlinear material as in Eq.~(\ref{selfkerr}) above, we can easily observe that those processes involving general CV states are still problematic as well as unsuccessful. Instead, we move our focus onto the Gaussian states for dealing with finite degrees of freedom only.  The Gaussian state is a quantum state having Wigner function of Gaussian shape or normal distribution. Since we can always displace the average value by a local displacement operation, so the only valuable information of the Gaussian states is lying on the second moment of canonical variables, which is generally called by covariance matrix (CvM). The CvM is positive and real symmetric $2n\times 2n$ matrix for any $n$-mode Gaussian state. Also Gaussian operation over the Gaussian states is defined as a unitary operation preserving Gaussian characteristics, which is homogeneous subgroup of Sp(2$n$,$\bb{R}$) for $n$-mode Gaussian state~\cite{Lupo14}.
 
There still exists a subtle problem of dimension-mode matching, however, even though we are dealing with finite degrees of freedom of Gaussian states. For the simplest example, let us consider a discrete quantum state of dimension 2 (i.e., qubit) and a single-mode Gaussian state. In this case, the number of free parameters for both cases are 3 under the consideration of the normalization. Next non-trivial example is a quantum state of dimension 4 (two-qubit cases) and a two-mode Gaussian state. Here, the number of free parameters of the two-qubit state is 15, but of the two-mode Gaussian state is only 10~\cite{Strang}. Consequently, we cannot make a simple correspondence between $d$-dimensional DV and $n$-mode Gaussian quantum states for general cases.

\section{Correspondence between MES and Gaussian MES} \label{CVDV}

As seen in Section~\ref{genco}, to find general corresponding relation between Gaussian states and DV quantum states is still challenging. Thus we need to move our concentration to the simpler case, that is, correspondence between maximally entangled states (MESs) in both regimes. Before describing our main results in detail, we need to briefly review what the Gaussian states and Gaussian operations are, and how the maximally mixed state (MMS)/MES can be defined in the Gaussian regime.

The EPR state can be well-matched to a MES of arbitrary dimension. In Gaussian regime, TMSV state is approaching to the EPR state in the limit of \emph{infinite} squeezing~\cite{NOPA}. Note that, on two-mode vacuum states, it is generally represented by
\begin{equation} \label{TMSV}
\ket{\psi(r)}_\tn{TMSV}=e^{\frac{r}{2}(\hat{a}\hat{b}-\hat{a}^\dag\hat{b}^\dag)}\ket{0}_A\ket{0}_B,
\end{equation} 
where $r$ is the squeezing parameter, and $\hat{a}$ and $\hat{b}$ the bosonic field operators. The state also can be derived from the quantum purification of the thermal state as $\ket{\psi(r)}_\tn{TMSV}=\sum_{n=0}^\infty\frac{(\tanh r)^n}{\cosh r}\ket{n}_A\ket{n}_B$ in Fock basis under the condition of the average photon number $\bar{n}=\sinh^2r$. Since the infinite squeezing is not physically plausible operation, thus we have to consider only finite squeezing. In other words, when we are dealing with Gaussian states physically, we need to confine our states in appropriate energy boundary. Then, we can guess a TMSV with finite squeezing parameter which might match to the MES of a specific dimension.  

The thermal state can be expressed in coherent state basis as $\rho_\tn{th}(\bar{n})=\frac{1}{\bar{n}\pi}\int e^{-\frac{|\alpha|^2}{\bar{n}}}|\alpha\rangle\!\langle\alpha| d^2\alpha$, where $\bar{n}$ is the average photon number. As $\bar{n}$ goes to infinity, it is obvious that the thermal state has uniform distribution with respect to the coherent state basis on the entire phase space; this fact means Gaussian maximally mixed state (GMMS). Instead, another version of GMMS is possible~\cite{Bradler} such that
\begin{align} \label{CVMMS}
\rho_{\tn{GMMS}}|_b&:=\frac{\1_b}{C_b}= \frac{1}{C}\int_b \proj{\alpha}{\alpha}d^2\alpha \nonumber\\
&=\frac{1}{b^2}\sum_{n=0}^\infty\left(1-\sum_{k=0}^n\frac{b^{2k}}{k!}e^{-b^2}\right)\proj{n}{n},
\end{align}
where $C=\pi b^2$ is the normalization constant. Since the coherent state $\ket{\alpha}$ is a Gaussian state, this GMMS is the convex sum of all Gaussian states  within radius $b$ from the origin of the phase space. This state indeed has the property of MMS in the sense that it can be used for private quantum channel and the same holds true for its general version including squeezing operations~\cite{JKL15}.

We can think a purification of $\rho_{\tn{GMMS}}|_b$ as
\begin{equation}
|\psi(b)\rangle_{\tn{GMES}}=\sum_{n=0}^{\infty}\sqrt{f(n,b)}|n\rangle _A |n\rangle _B,
\end{equation}
where the coefficient is explicitly given by $f(n,b)=(1-\sum_{k=0}^n\frac{b^{2k}}{k!}e^{-b^2})/b^2$~\cite{JeongLim}. Like the TMSV state, this state is indeed another candidate for GMES as $b \rightarrow \infty$. Also we infer that it can be related with MES of certain dimension when $b$ is finite. Now we can raise a reasonable question. Which candidate is more appropriate for the concept of GMES? We can answer this question quantitatively in the next section, our main results.

\begin{figure}[!t]
\includegraphics[width=9.3cm]{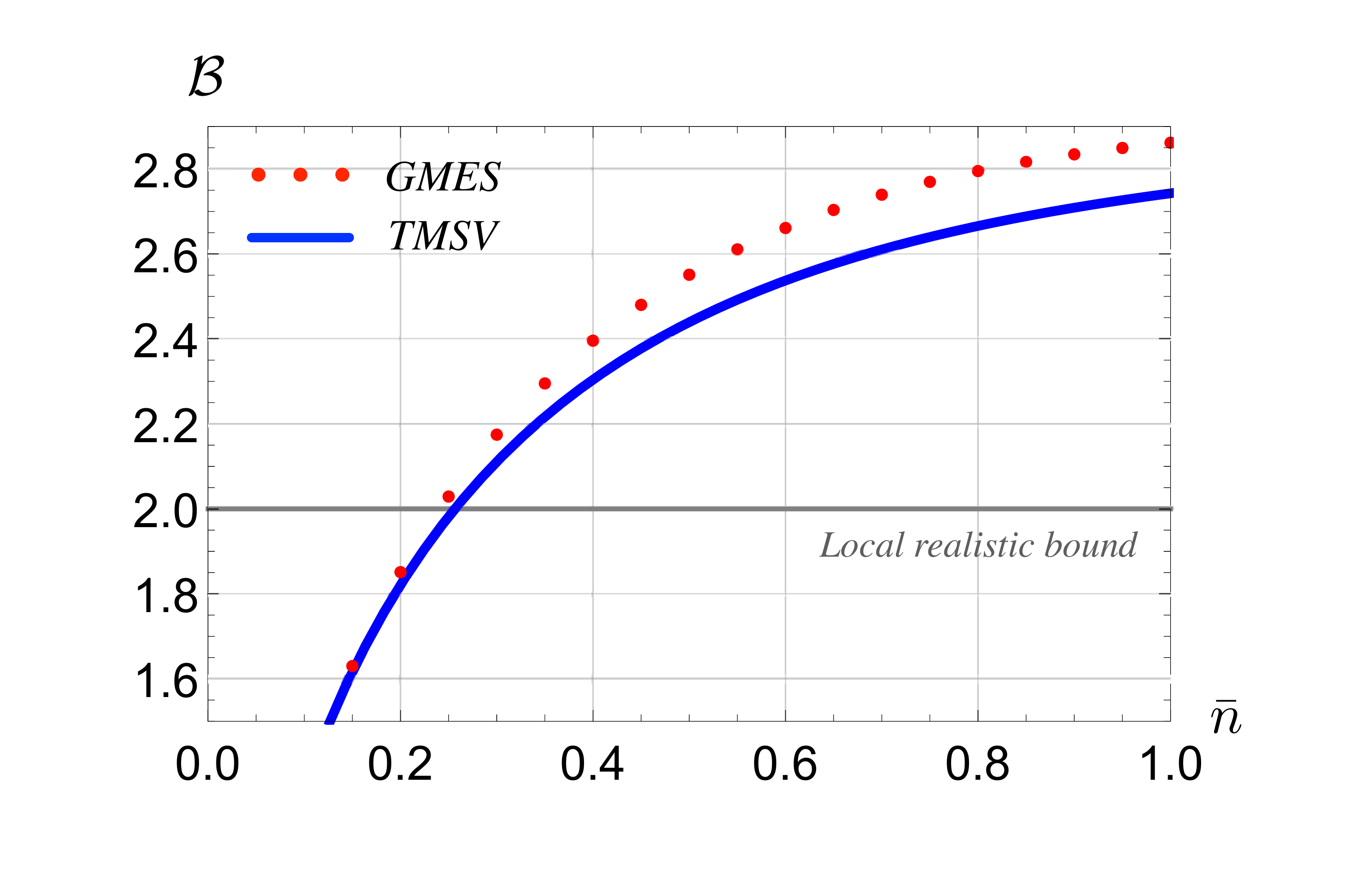}
\caption{Maximal Bell functions of the two-qutrit for our GMES and TMSV state under the average photon number $\bar{n}$. The maximal Bell function of the GMES is larger than of TMSV state after reaching around the local realistic bound 2. }
\label{fig1}
\end{figure}

\section{Main results} \label{main}

We first introduce a Bell's test on two-qutrit systems, and investigate the violation of Bell's inequality via comparison of two specific candidates having quantum entanglement. For the MES of bipartite qudit system, the amount of violation for Bell's inequality is already known~\cite{CGLMP}. However, the coefficients of GMESs we proposed are not uniform under the condition of finite squeezing and at a given boundary, so we cannot directly apply this method to our case. Although, the analytic formula of maximal Bell function is known, for less general case, i.e., two-qutrit~\cite{qutrit}, as
\begin{equation} \label{qutrits}
\ket{\psi}=\sum_{k=0}^2 a_k \ket{k}_A\ket{k}_B,
\end{equation}
where $a_k$'s are real coefficients and $\forall k, \ket{k}_{A(B)}$ denote the orthonormal basis for the qutrit on the system $A(B)$, respectively. For the two-qutrit case, the Bell-CHSH inequality can be expressed as~\cite{CHSH1,CHSH2}
\begin{equation}\label{CHSH}
-4\leq \cl{B} \leq 2.
\end{equation}
Also, the Bell function or operator $\cl{B}$ is written as in the form of
\begin{align} \label{Bell}
\cl{B}:=&\tn{Re}\left[Q_{11}+Q_{12}-Q_{21}+Q_{22}\right] \nonumber\\
&+\tfrac{1}{\sqrt{3}}\tn{Im}\left[Q_{11}-Q_{12}-Q_{21}+Q_{22}\right],
\end{align}
where $Q_{ij}$ is the correlation function between Alice's and Bob's measurements for two observables ($\forall i,j\in\{1,2\}$).
Then, a maximal value of the Bell function, Eq.~(\ref{Bell}), is known as $\cl{B}_{\max}=4|a_0a_1|+4/\sqrt{3}(|a_0a_2|+|a_1a_2|)$~\cite{qutrit}, if $\max\{ |a_0|,|a_1|,|a_2|\} \leq \sqrt{18+9\sqrt{3}}/2$, which is in our case we want. We have $a_k={(\tanh{r})^k / \sqrt{1+(\tanh{r})^2+(\tanh{r})^4}}~$ for the TMSV state, and
\begin{equation}
a_k=\frac{\sqrt{1-\frac{\Gamma \left(k+1,b^2\right)}{\Gamma (k+1)}}}{\sqrt{-e^{-b^2}-\Gamma \left(2,b^2\right)-\frac{\Gamma \left(3,b^2\right)}{2}+3}}(:=b_k)
\end{equation}
for our GMES, where $\Gamma(n+1,b^2)=\int_{b^2}^\infty t^n e^{-t}dt$ is the incomplete gamma function. (Note that, for convenience, $b_k$ emphasizes the boundary radius $b$ in the Gaussian state.) In Fig.~\ref{fig1}, it is shown that our GMES has higher value for the maximal Bell function than that of TMSV state after passing the local realistic bound, though the difference is not so much. Note that in the limit of $\bar{n}\rightarrow \infty$, then the value of maximal Bell function is $ \sim$2.873---it was known for maximally entangled two-qutrit states. For the low range of the average photon number ($\bar{n}$), our GMES has the maximal entanglement more effectively, because the average photon number is one kind of physical resources for generating a quantum state in real experiment. 

\begin{figure}[!t]
\includegraphics[width=9.2cm]{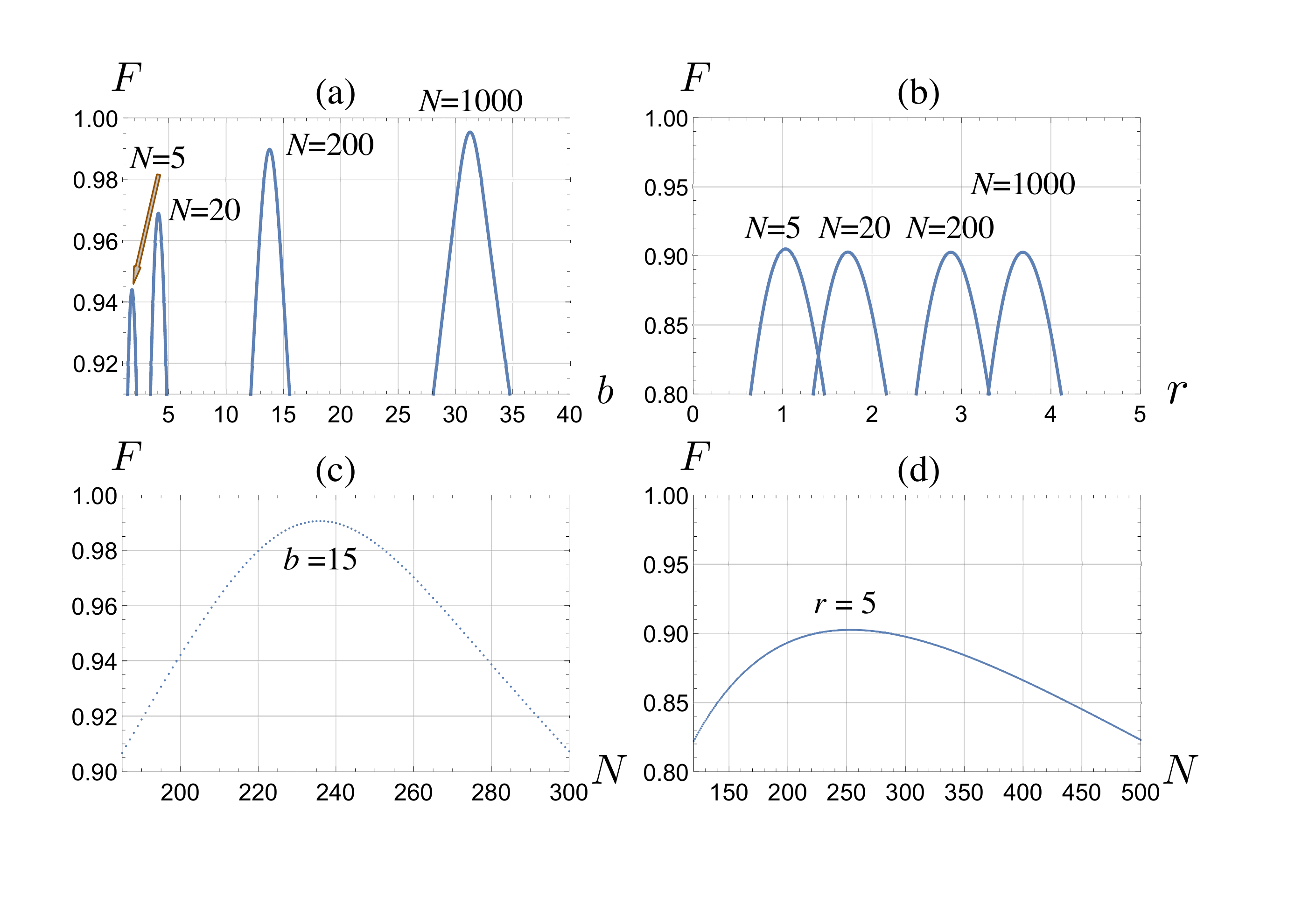}
\caption{(a), (b) Fidelities between GMESs/TMSV states and MESs of $N \times N$ dimension. Fidelities of GMESs are much higher than of TMSV states for those $N=5,20,200$, and 1000. (c) Fidelity between various MESs and GMES of $b=15$. The maximum fidelity is around 0.99. (d) Fidelity between various MESs and TMSV state of $r=5$. Also, the maximum fidelity is achieved around 0.9.}
\label{fig2}
\end{figure}

For our original purpose, however, we need more direct comparison for the maximally entangled states. In this reason, we will check which dimension of MES corresponds to the TMSV state or the GMES with finite parameters by calculating a fidelity between those states. The fidelity between two pure states $\ket{\psi}$ and $\ket{\phi}$ is generally defined by $F:=| \langle \psi | \phi \rangle |$, thus we easily calculate and plot these results as in Fig.~\ref{fig2}. Note that in the limit $N \rightarrow \infty$, fidelities converges unity for both cases. However, our GMES is much more practical in a sense that the fidelity is already over 0.99 for $N=200$. On the contrary, for the case of TMSV state, the fidelity is around 0.9 even for very high $N$. In other words, if we want to make a kind of Gaussian state which corresponds to $N\times N$-maximally entangled state, our GMES is more appropriate than TMSV state. For the converse cases, as in Fig.~\ref{fig2} (c) and (d), also we obtain high fidelities when we use the GMES of specific $b$. It can be understood that our GMES comes from the uniform distribution Eq.~(\ref{CVMMS}) but TMSV state from the thermal state whose distribution is Gaussian. Since our GMMS already has ``infinite temperature'' within a boundary $b$, the purified state in GMES has more uniform distribution that corresponds to the discrete-variable MES.

Experimentally, it is well-known that TMSV state can be generated for relatively small squeezing parameter~\cite{TMSV}. Although we don't have concrete idea for generating our GMES, but we have for GMMS. An input coherent state passes through a phase shifter and a beam splitter, and assume we don't know phase and transmissivity. Then a final state has completely unknown phase and amplitude, the only boundary $b$ can be estimated from the input amplitude. This is the exact GMMS as in Eq.~(\ref{CVMMS}).

\section{Discussion} \label{discussion}

We showed that the two-mode Gaussian maximally entangled state (GMES) we proposed has more proper correspondence with discrete MES than the TMSV state. There might be many kinds of Gaussian maximally entangled states in contrast with discrete cases. Our GMES is might be \emph{optimal}, because it is from the uniform distribution Eq.~(\ref{CVMMS}) although more consideration is needed for its rigorousness. 

Unfortunately, there are several problems we have. First of all, we don't know yet the method for experimental realization of the GMES. It can be obtained from the purification of GMMS or be from another method. Second, we only investigated the corresponding relation between MMS/MES in the Gaussian and discrete regimes. Especially, only two-mode state is enough for this MES-GMES correspondence, beside we should consider multi-mode Gaussian states for general cases. Multi-mode Gaussian states have rich structure like genuine multi-partite entanglement, but also have very different structure even for three-mode cases~\cite{multi}. 

Another interesting problems can be found from bound entangled states, which are entangled state but cannot be distillable. For the discrete case, the minimal dimension of the bound entangled state is $2 \otimes 4$ or $3 \otimes 3$. On the contrary, $2 \oplus 2$ modes is minimum for Gaussian states, but there is no bound entanglement for $1 \oplus n$ modes~\cite{bound}. This implies that the mode-dimension correspondence problem is still open, and we will study bound entangled states in both regimes for the next simplest case.

\section*{ACKNOWLEDGMENTS}

This work was supported by Basic Science Research Program through the National Research Foundation of Korea (NRF) funded by the Ministry of Education (NRF-2017R1A6A3A01007264) and the Ministry of Science and ICT (NRF-2016R1A2B4014928). J.K. appreciates the financial support by the KIST Institutional Program (Project No. 2E26680-16-P025). K.J. acknowledges financial support by the National Research Foundation of Korea (NRF) through a grant funded by the Korean government (Ministry of Science and ICT) (NRF-2017R1E1A1A03070510 \& NRF-2017R1A5A1015626).

\end{document}